\documentclass[]{interact}

\usepackage{epstopdf,float}

\usepackage[numbers,sort&compress]{natbib}
\usepackage{bm,xcolor,lscape}

\usepackage{epsfig,array,bbm,fancyhdr}
\usepackage{subfigure,multirow,amsthm,subfig}
\usepackage{amsmath,latexsym,textcomp}
\usepackage{setspace,mathrsfs,amssymb}
\usepackage{bm,xcolor,lscape,graphicx}
\usepackage{booktabs,float,epic,natbib}
\usepackage{comment,txfonts,mathtools,url,caption} 

\usepackage{graphicx}
\usepackage{subcaption}

\usepackage{listings,numprint,enumerate,bbm}
\usepackage{verbatim,dsfont,scalerel}

\bibpunct[, ]{[}{]}{,}{n}{,}{,}

\renewcommand\bibfont{\fontsize{10}{12}\selectfont}
\makeatletter
\def\NAT@def@citea{\def\@citea{\NAT@separator}}
\makeatother

\theoremstyle{plain}

\theoremstyle{definition}

\theoremstyle{remark}

\begin{document}


\title{Doubly Robust Weighted Regression for Causal Inference under Confounder Missingness}

\author{
\name{Md Shaddam Hossain Bagmar\thanks{CONTACT Md. Shaddam Hossain Bagmar. Email: shaddam.bagmar@gmail.com} and Hua Shen}
\affil{Department of Mathematics and Statistics, University of Calgary, Calgary, Alberta, Canada}
}


\maketitle

\begin{abstract}
 Missingness in confounders are common in observational studies and present substantial challenges for causal effect estimation by weakening identification and increasing sensitivity to model misspecification. Within the missing-indicator framework, existing approaches typically rely on a single working model and achieve consistency only when that model is correctly specified, and are therefore singly robust. In this article, we develop a doubly robust missing indicator weighted ordinary least squares (MI-WOLS) estimator with partially observed confounders. The proposed method integrates propensity score–based weighting into outcome regression under the missing-indicator data representation, yielding a class of balancing weights that account for both confounders and their missingness indicators. Under the missingness-strongly-ignorable treatment allocation assumption and assuming either a Conditionally Independent Treatment or Conditionally Independent Outcome structure, the MI-WOLS estimator is consistent when at least the treatment or the outcome model is correctly specified. Simulation studies support the theoretical robustness of the MI-WOLS estimator, demonstrating negligible bias, accurate sandwich-based variance estimation, and near-nominal coverage probability across a wide range of data-generating scenarios. An illustrative application using a simulated data designed to reflect a real-world kidney function study further demonstrates the interpretability and practical feasibility of the method, offering a flexible, doubly robust alternative to existing singly robust estimators.

\end{abstract}


\section{Introduction}

Understanding causal relationships is central to modern health research, where investigators aim to estimate treatment effects while accounting for differences in patient characteristics; however, classical methods based on associational contrasts may fail in the presence of confounding \citep{robins2000marginal,weisberg2011bias,glass2013causal}. Causal inference methods offer a principled framework for reducing such bias by modelling either the treatment assignment mechanism or the outcome, but valid estimation depends on correctly specified models and on assumptions that justify fair comparisons between treatment groups \citep{rosenbaum1983central,weisberg2011bias,hernan2019causal} —assumptions that are especially crucial in clinical settings where treatment decisions depend on patient history, comorbidities, and laboratory data \citep{tilden2018causal,weaver2024necessity}. A major practical challenge in these analyses is missing data, particularly missing confounders, which are common in clinical research because of irregular follow-up, noncompliance, administrative factors, or other unobserved reasons. Complete-case analysis yields unbiased estimates only under the restrictive missing completely at random (MCAR) assumption and may provide biased and inefficient results when missingness follows a missing at random (MAR) or missing not at random (MNAR) mechanism \citep{li2013weighting}. Consequently, the presence of missing confounders can weaken the `no unmeasured confounding' assumption and compromise causal identification, even when treatment and outcome variables are fully observed \citep{daniels2023bayesian}.

Several recent studies have addressed confounder missingness in causal inference. \citet{blake2020propensity,blake2020estimating} introduced a missing-indicator-based propensity score estimator under MAR and later developed a corresponding outcome regression estimator, both of which rely on the missingness-strongly-ignorable treatment allocation (mSITA) assumption together with either a Conditionally Independent Treatment (CIT) or Conditionally Independent Outcome (CIO) assumption. Because each method depends on the correct specification of a single working model—the treatment model for the propensity score estimator or the outcome model for the regression estimator—these estimators are singly robust and provide biased estimates when their respective models are misspecified. \citet{bagmar2022causal} and \citet{mayer2020doubly} examined doubly robust augmented inverse probability weighting (AIPW) estimators in settings with missing confounders, with a primary focus on estimating average treatment effects. While these approaches provide insight into whether a treatment is beneficial at the population level, they do not directly support individualized treatment decision-making. In parallel, the dynamic treatment regime (DTR) literature \citep{chakraborty2013statistical,chakraborty2014dynamic} has introduced a doubly robust weighted regression approach for estimating optimal treatment rules at the patient level \citep{nadi2025recent}. \citet{wallace2015doubly} first introduced dynamic weighted ordinary least squares (dWOLS) for estimating optimal treatment regimes with continuous outcomes and binary treatments. The dWOLS method employs covariate-balancing weights to obtain consistent estimates of blip functions when either the treatment or the outcome model is correctly specified. Subsequent work has extended the dWOLS framework to accommodate continuous treatments \citep{schulz2021doubly}, binary outcomes \citep{jiang2022doubly,bian2024variable}, survival outcomes \citep{simoneau2020estimating,zhang2022doubly}, and ordinal outcomes in the presence of interference \citep{jiang2024estimating}. Despite these methodological advances, existing dWOLS approaches generally rely on fully observed confounders and do not explicitly address missing confounder data.

The present work integrates and extends two areas of the causal inference literature: balancing-weight approaches for treatment effect estimation \citep{wallace2015doubly} and missing-indicator methods for handling partially observed confounders \citep{blake2020propensity,blake2020estimating}. Most existing doubly robust estimators generally assume fully observed confounders, whereas methods for partially observed confounders under the missing-indicator framework are typically singly robust. To address this gap, we develop a unified missing-indicator weighted ordinary least squares (MI-WOLS) estimator that defines propensity score–based balancing weights on an augmented covariate space that includes both confounders and their missingness indicators. This extension preserves the covariate-balancing property while allowing for confounder missingness under a MAR mechanism, yielding a doubly robust estimator that remains consistent when either the treatment or the outcome model is correctly specified under mSITA, together with either CIT or CIO. Throughout the paper, the terms \textit{confounder missingness} and \textit{partially observed confounders} are used interchangeably to describe confounders observed for some individuals but missing for others. These terms do not refer to unmeasured confounders, which are unavailable for all individuals.

The remainder of the article is structured as follows. Section \ref{NOTATION} presents the causal framework, assumptions, and notation. Section \ref{METHOD} introduces the MI-WOLS estimator and defines the balancing condition under confounder missingness. Section \ref{SIMULATION} presents simulation results evaluating the finite sample performance of the proposed doubly robust MI-WOLS estimator and compares it with two established doubly robust causal methods, AIPW and G-estimation. Section \ref{ILLUSTRATION} provides an illustrative application to the effect of angiotensin-converting enzyme inhibitors or angiotensin receptor blockers (ACEI/ARBs) treatment on estimated glomerular filtration rate (eGFR). Section \ref{DISCUSSION} concludes with a discussion and directions for future work.

\section{Causal framework, identification assumptions, and notation}
\label{NOTATION}
In this section, we introduce the notation, potential outcomes formulation, and the key assumptions required for the identification of causal effects when confounders may be partially observed. Our presentation follows the general structure developed by \citet{blake2020propensity,blake2020estimating}, who studied singly robust estimators under a missing–indicator approach, and extends those ideas to the weighted regression setting considered in this work.

\subsection{Notation and setup}

Let $C$ be the vector of fully observed confounders, $Z \in \{0,1\}$ denote a binary treatment, and $Y$ a continuous outcome. Let $X$ represent a vector of confounders that are subject to missingness, and let $R$ denote the associated missingness indicators, where $R=1$ indicates that $X$ is observed and $R=0$ otherwise. When needed, we write $X=(X^{\mathrm{obs}}, X^{\mathrm{mis}})$ to distinguish the observed and missing components.

We adopt the standard potential outcomes framework. For $z\in\{0,1\}$, let $Y(z)$ denote the potential outcome that would be observed if treatment were set to $z$. We assume consistency, i.e.\ $Y = Y(1)Z + Y(0)(1-Z)$; and positivity, meaning that for any possible covariate pattern, the probability of receiving either treatment level is strictly between $0$ and $1$. Throughout, we operate under an MAR structure, so that the missingness indicators may depend on observed data but not on the unobserved values $X^{\mathrm{mis}}$ given $(C, X^{\mathrm{obs}})$.

\subsection{Identification under confounder missingness: mSITA, CIT, and CIO}
\label{ASSUMPTION}
In settings where some confounders are partially observed, identification of causal effects requires additional conditions relating the treatment assignment mechanism, the outcome process, and the missingness indicators. The missing–indicator strategy replaces each partially observed confounder with its observed value and an accompanying missingness indicator, and valid inference relies on a set of assumptions that ensure these variables adequately adjust for confounding. We outline below the assumptions needed for identifying the average treatment effect in our setting.

We first require that the treatment assignment is unconfounded given the full set of covariates and their missingness indicators. Formally, for each $z \in \{0,1\}$,
\[
Z \,\perp\!\!\!\perp\, Y(z) \mid C, X, R,
\]
which is commonly referred to as the missingness-strongly ignorable treatment assignment (mSITA) assumption. This condition extends the usual no–unmeasured–confounding assumption to the situation where the treatment mechanism may vary across strata defined by the missing components of $X$. In practical terms, this assumption states that, after accounting for all observed covariates and the missingness patterns, there are no unmeasured factors that jointly influence treatment assignment and potential outcomes. In particular, it allows the treatment mechanism to differ across missingness strata, but requires that within each stratum defined by $\left(C, X, R\right)$, treatment assignment is effectively as good as random. This assumption is plausible in settings where the available covariates and missingness indicators capture the key clinical or administrative factors driving both treatment decisions and outcomes.

Because $X^{\mathrm{mis}}$ is not observed, mSITA alone is insufficient for the causal effect estimation. Causal effect identification proceeds by assuming that the unobserved portion of $X$ is irrelevant either for the treatment mechanism or for the outcome, conditional on the observed data, to ensure that the missing components do not introduce residual confounding. These conditions are formalized through the following two alternative assumptions.

The Conditionally Independent Treatment (CIT) assumption requires that
\[
Z \,\perp\!\!\!\perp\, X^{\mathrm{mis}}
   \mid C, X^{\mathrm{obs}}, R,
\]
which implies that, conditional on the observed covariates and missingness indicators, treatment assignment does not depend on the unobserved components of the confounders. In practice, this assumption is plausible when treatment decisions are based primarily on recorded or routinely collected information, and any missing covariates do not play a direct role in determining treatment.

Alternatively, the Conditionally Independent Outcome (CIO) assumption states that for each
$z \in \{0,1\}$,
\[
Y(z) \,\perp\!\!\!\perp\, X^{\mathrm{mis}}
   \mid C, X^{\mathrm{obs}}, R,
\]
meaning that the unobserved components of the confounders do not provide additional predictive information about the potential outcomes beyond what is captured by the observed data. This assumption may be reasonable when the missing covariates are weak predictors of the outcome or when their effects are adequately proxied by observed variables.

Together, mSITA with either CIT or CIO ensures that the missing-indicator representation suffices to control for confounding, allowing causal effects to be identified despite partial observability of the confounders. For the singly robust estimators of \citet{blake2020propensity,blake2020estimating}, consistent estimation also requires correct specification of the associated working model: the treatment model for propensity score-based estimation or the outcome model for outcome regression estimation. In contrast, the proposed MI-WOLS estimator achieves double robustness. Under mSITA and either CIT or CIO, consistent estimation of the causal effect is obtained provided that at least one of the two working models-the treatment model defining the propensity weights or the outcome model-is correctly specified. Thus, our estimator remains valid even when one of the models is misspecified.

\subsection{Implications and diagrammatic illustration}
\begin{figure}[H]%
    \centering
\includegraphics[width=8cm,height=5cm]{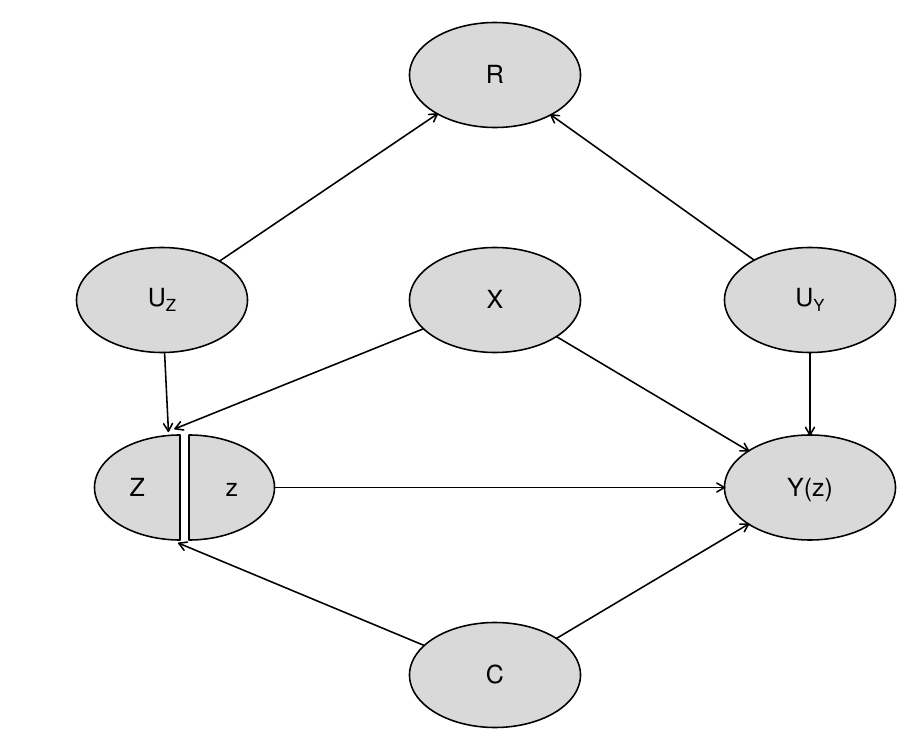} %
    \caption{Least-restrictive DAG under an MAR mechanism, depicting the causal structure among treatment $Z$, potential outcomes $Y(z)$, fully observed covariates $C$, partially observed confounders $X$, missingness indicators $R$, and latent variables $U_Z$ and $U_Y$.}%
    \label{DAGS}%
\end{figure}

We use directed acyclic graphs (DAGs) to illustrate the identifying assumptions introduced in Section~\ref{ASSUMPTION} and to clarify their causal implications. Figure~\ref{DAGS} presents a least-restrictive DAG that includes all plausible causal relationships among the treatment $Z$, potential outcomes $Y(z)$, fully observed confounders $C$, partially observed confounders $X$, missingness indicators $R$, and latent variables $U_Z$ and $U_Y$. Throughout, we assume an MAR mechanism, represented graphically by the absence of a direct arrow from the partially observed confounders $X$ to its missingness indicator $R$.

Under this general DAG, the mSITA assumption does not hold. Conditioning on $(C, X, R)$ fails to block the backdoor path
\begin{align*}
Z \leftarrow U_Z \rightarrow R \leftarrow U_Y \rightarrow Y(z),   
\end{align*}
so treatment assignment and potential outcomes remain associated. Graphically, mSITA is satisfied only when this path is blocked, which occurs if at least one of the arrows linking $U_Z$ to $Z$, $U_Z$ to $R$, $U_Y$ to $R$, or $U_Y$ to $Y(z)$ is absent. DAGs corresponding to these restricted cases are provided in Figure S1 of the Supplementary Materials.

The CIT and CIO assumptions characterize how a partially observed confounder operates when it is missing. Under the CIT assumption, the missing component of $X$ does not influence treatment assignment; specifically, for individuals with $R = 0$, there is no causal path from $X$ to $Z$. Under the CIO assumption, the missing component of $X$ does not affect the potential outcomes, implying that no causal path exists from $X$ to $Y(z)$ when $R = 0$. When both CIT and CIO hold, $X$ is causally disconnected from both $Z$ and $Y(z)$ within the missingness stratum. The corresponding causal structures are illustrated using DAGs in Figure~S2 of the Supplementary Materials.

\section{The MI-WOLS estimator and its balancing property}
\label{METHOD}
Weighted regression methods have played a central role in the estimation of causal effects in  DTR settings. In particular, the dWOLS estimator of \citet{wallace2015doubly} uses propensity score based absolute weights to construct a pseudo-population where treatment is independent of measured covariates, thereby yielding consistent estimation of blip parameters even when either the treatment or the outcome model is misspecified. This methodology was later extended to continuous treatments by \citet{schulz2021doubly}, who formalized a general balancing property that guarantees consistent weighted-regression estimation.

In this section, we adapt these ideas to settings in which some confounders are partially observed. Specifically, we generalize the balancing property of the dWOLS weights to accommodate the missing-indicator representation of \citet{blake2020propensity,blake2020estimating}, thereby enabling doubly robust estimation when either the treatment or the outcome model is correctly specified. We first introduce the weighted estimating equations used in our proposed MI-WOLS estimator, and then formalize the balancing condition required for consistency under confounder missingness.

\subsection{The MI-WOLS estimating equations}

Let $V=(C, X)$ denote the full set of confounders, where $C$ is fully observed, and $X$ is partially observed. Let $R$ denote the corresponding missingness indicators and define $H = (V, R)$ as the covariate vector under the missing-indicator data representation. For continuous covariates, missing values are replaced by a fixed constant, and both the modified covariate and its indicator are included in the model, while for categorical covariates, an additional `missing' category is introduced \citep{pedersen2017missing}. We specify the following treatment and outcome models:
\begin{align}
P(Z=1 \mid H) &= \pi(H;\alpha) = expit\left(H^{\alpha} \bm{\alpha}\right), 
\nonumber\\
E\{Y \mid Z,H\} &= H^{\beta} \bm{\beta} + H^{\psi}Z\bm{\psi},    
\end{align}
where $expit(a) = 1/(1+exp(-a))$, $H^{\alpha}$ is the design matrix for the treatment model, $\alpha$ is the vector of regression parameters of the treatment model, $H^{\beta}$ and $H^{\psi}$ are the design matrices for the treatment free part and blip part of the outcome model, respectively. Here, all the design matrices $H^{\alpha}, H^{\beta}$ and $H^{\psi}$ are appropriate functional forms derived from $H$ and also include a column of 1 to accommodate the intercept term of the model whereas for the blip design matrix $H^{\psi}$ the column of 1 will allow the main effects of the treatment. Lastly, $\beta$ and $\psi$ are the vectors of regression parameters of the treatment-free part and blip part of the outcome model. The primary interest of precision medicine is to consistently estimate the blip model parameter $\psi$.

Following the weighted-least-squares formulation in \citet{wallace2015doubly}, the proposed MI-WOLS estimator solves
\begin{equation}
0
=
\sum_{i=1}^n
\begin{pmatrix}
H^{\beta}_i \\
Z_i H^{\psi}_i
\end{pmatrix}
w_i(z,h)
\Bigl\{
Y_i - H^{\beta}_i \bm{\beta} - H_i^{\psi}Z_i\bm{\psi}
\Bigr\},
\label{eq:WOLS}
\end{equation}
where $w_i(z,h)$ is a weight constructed from the propensity score $\pi(H_i;\hat{\alpha})$. Different choices of $w_i$ correspond to different weighting schemes, all of which are admissible provided they satisfy the balancing property stated below.

\subsection{Balancing weights under the missing-indicator representation}

In the complete-data setting, where all confounders \(V\) are fully observed, \citet{wallace2015doubly} showed that consistency of the weighted regression estimator is guaranteed by the existence of a weight function \(w(z,v)\) that satisfies the following covariate balancing condition:

\[
\pi(v)\, w(1,v) = \{1-\pi(v)\} w(0,v),
\]
ensuring that the weighted pseudo-population satisfies independence between treatment and covariates. \citet{schulz2021doubly} later demonstrated that this property extends to continuous treatments under an analogous identity involving the generalized propensity score.

Under the missing indicator approach, where $H = (V, R)$ is the augmented covariate vector, the corresponding balancing condition naturally extends to hold with respect to $H$ as:
\begin{align}
\pi(h)\, w(1,h)
&=
(1-\pi(h))\, w(0,h) \;\;\text{or equivalently}\nonumber\\
\pi(v,r)\, w(1,v,r)
&=
(1-\pi(v,r))\, w(0,v,r)
\label{eq:balance-missing}
\end{align}
This condition ensures that after weighting, treatment assignment is independent of observed covariates (including missingness indicators), mimicking a randomized experiment. We restate the balancing property formally as below:\\

\noindent{\bf Theorem 3.1 (Balancing property under missing indicator approach).}
\emph{
Suppose that the missingness-strongly-ignorable treatment assignment (mSITA) assumption holds, together with either the Conditionally Independent Treatment (CIT) assumption or the Conditionally Independent Outcome (CIO) assumption. If either (i) the treatment model $P(Z=1 \mid H)$ or (ii) the outcome model $E(Y\mid Z, H)$ is correctly specified, then the MI-WOLS estimator obtained by solving \eqref{eq:WOLS} is consistent for the blip parameter $\psi$ provided that the weight function satisfies the balancing condition 
\eqref{eq:balance-missing}.
}
\\

The proof follows the arguments of \citet{wallace2015doubly} and \citet{schulz2021doubly}, with additional conditioning on $(X, R)$ and decomposition of the pseudo-population density under the missing-indicator representation. Because these derivations are algebraically lengthy, complete details are provided in Section S3 of the Supplementary Materials.

\section{Simulation studies}
\label{SIMULATION}
In this section, we investigate the finite-sample properties of the MI-WOLS estimator across a range of confounder-missingness settings. Our simulation framework extends the data-generating mechanisms of \citet{blake2020estimating}, originally designed to assess singly robust estimators, by introducing a broader set of scenarios that enable systematic comparison between our proposed MI-WOLS estimator and competing approaches.

\subsection{Data-generating mechanisms and design}\label{SUBSECSIM}
The simulation framework is designed to mimic a realistic causal structure in which a partially observed confounder affects both treatment assignment and the outcome through separate mechanisms. Let $X$ be a binary confounder that may be missing, $C$ a fully observed binary confounder, $R$ the indicator of whether $X$ is observed, $Z$ a binary treatment, and $Y$ a continuous outcome. In addition, let $U_Z$ and $U_Y$ denote independent standard normal variables that act as latent common causes for $(Z, R)$ and $(Y, R)$, respectively. To systematically explore violations of key assumptions and model specifications, a set of parameters is varied across structured scenarios. The resulting data-generating mechanisms are defined by the following distributions:
\begin{align} & X \sim Binomial(1, 0.67);\;\;C \sim Binomial(1, 0.58)\\ \nonumber & R\sim Binomial(1, 1-expit(-0.5+1.48U_Z + 1.36 U_Y))\\ \nonumber Z\sim Binomial(&1, expit(-1.2 +\tau U_Z + 1.38 XR + \lambda X(1-R) + 2R + 1.69C + \delta_Z CR))\\ Y \sim Normal(&1 + \psi_0 Z - 2.2\tau U_Y - 1.55 XR + \gamma X(1-R) + 1.8R - 1.7C + \delta_Y CR, 3)\nonumber \end{align}
To examine performance across a range of causal structures, the parameters 
$\tau \in \{0,1.25\}$, $\lambda \in \{0,1.38\}$, $\gamma \in \{0,-1.55\}$, $\delta_Z \in \{0,-4.2\}$, and $\delta_Y \in \{0,-4.2\}$
were varied such that setting $\tau = 0$ satisfied the mSITA assumption, while choosing $\lambda = 0$, $\gamma = 0$, $\delta_Z = 0$, or $\delta_Y = 0$ ensured the CIT assumption, the CIO assumption, correct specification of the treatment model, or correct specification of the outcome model, respectively. Under the missing indicator approach, correct model specification assumes no interaction between the fully observed confounder and the missing indicator, implying that the confounder's effect on treatment assignment and the outcome is the same across missingness patterns. The target estimand was the marginal treatment effect $\psi_0$, fixed at $-2.35$ to represent the average treatment effect under homogeneous effects. Across the different data-generating scenarios, the proportion of treated individuals ranged from 40.6\% to 86.2\%, while the percentage of missing values for the confounder $X$ varied between 32.6\% and 52.2\%.

To evaluate estimator performance under different assumption structures, we considered four model-specification scenarios defined by the treatment and outcome models: both correctly specified (CC), only the treatment model correctly specified (CI), only the outcome model correctly specified (IC), and both misspecified (II). For each scenario, we generated datasets of size $n = 500$ and repeated the simulation 1000 times. We compared several weighting approaches \citep{schulz2021doubly}, including absolute value weights (ABS), inverse probability weights (IPW), stabilized inverse probability weights (SIPW), and an unweighted estimator (UNW) which corresponds to Blake’s singly robust outcome regression estimator \citep{blake2020estimating}. By construction, the ABS and IPW weighting methods satisfy the covariate balancing condition defined in Equation (\ref{eq:balance-missing}), whereas SIPW does not generally achieve exact covariate balance \citep{schulz2021doubly}. Nevertheless, SIPW remains a consistent estimator under the standard identification assumptions required for IPW \citep{robins2000marginals,schulz2021doubly}. By incorporating the marginal treatment probability into the IPW weighting scheme, SIPW stabilizes the weights, thereby reducing the influence of extreme weights and improving estimation efficiency. As a result, despite not achieving exact covariate balance, SIPW yields consistent treatment effect estimates under correct specification of the treatment model, a property that is reflected in its empirical performance across the simulation studies. Formal definitions of the weight functions are provided in Supplementary Materials, Section S3.2. 

\subsection{Double robustness under model misspecification}

This subsection evaluates the double robustness of the MI-WOLS estimator in settings with missing confounders. Using the simulation design from Section \ref{SUBSECSIM}, we examine whether the estimator remains consistent when either the treatment or the outcome model is correctly specified. The results highlight how the MI-WOLS approach improves on existing singly robust methods by recovering the true causal effect under partial model misspecification.

 \begin{figure}[H]
 \centering
  \includegraphics[width=1\textwidth]{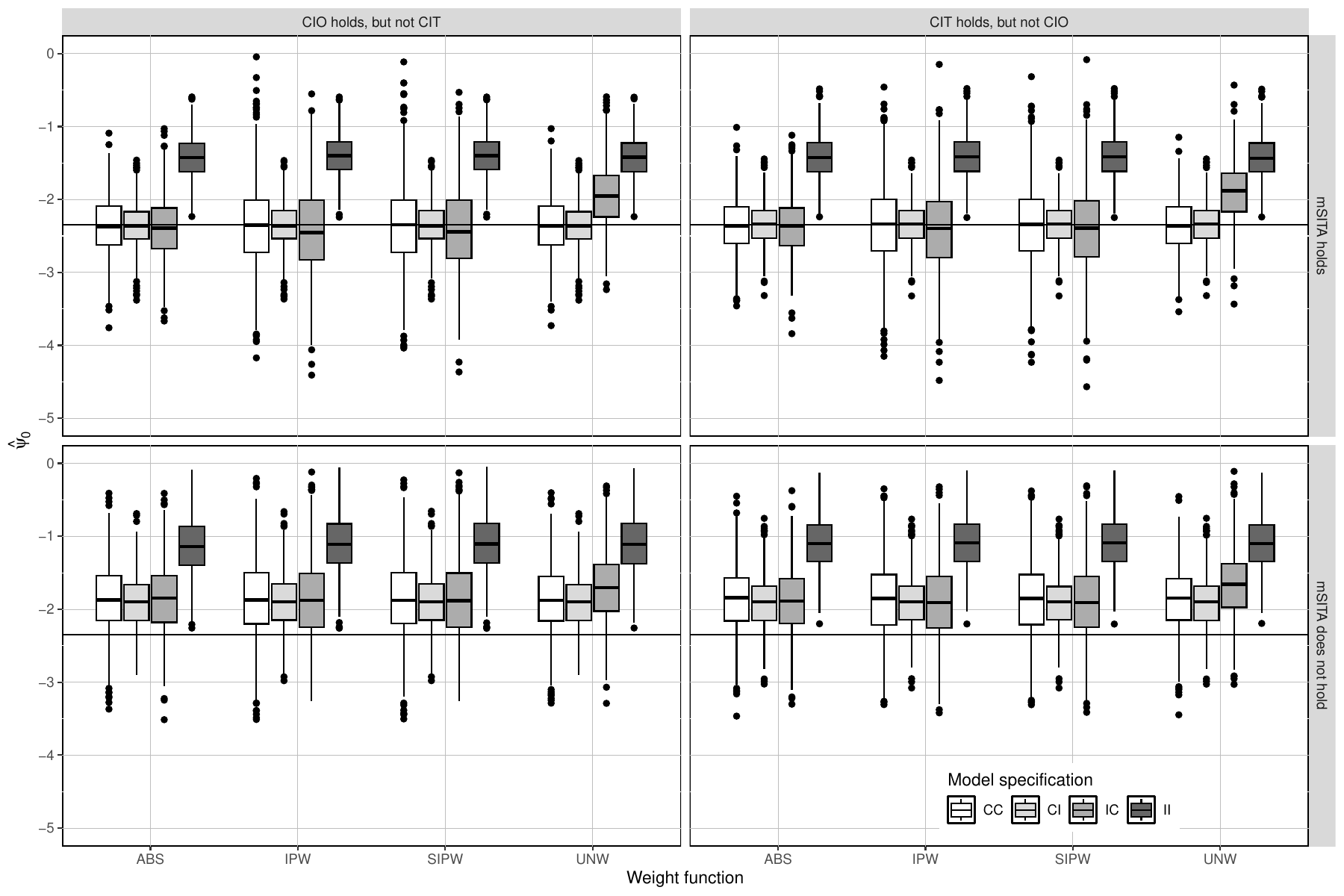}   
 \caption{Box plots of estimated treatment effects from the MI-WOLS estimator under scenarios where exactly one of the CIT or CIO assumptions hold. Results are shown separately for settings in which the mSITA assumption holds and does not hold, with a true effect of $-2.35$.}
 \label{BOX1}
 \end{figure}

\begin{figure}[H]
 \centering
  \includegraphics[width=1\textwidth]{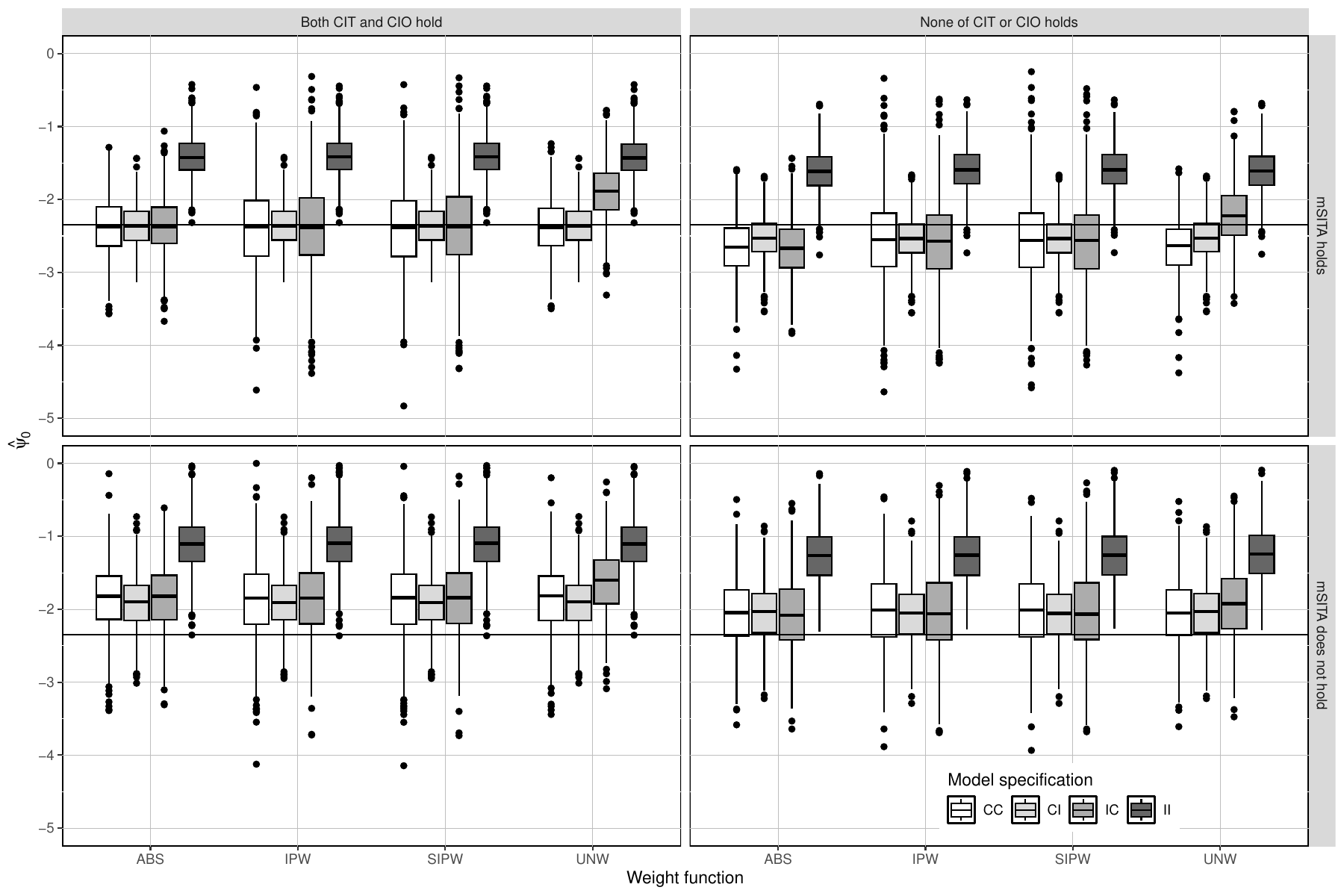}
 \caption{Box plots of estimated treatment effects from the MI-WOLS estimator under extreme identification scenarios where both CIT and CIO hold or neither holds. Results are shown separately for settings in which the mSITA assumption holds and does not hold, with a true effect of $-2.35$.}
 \label{BOX2}
 \end{figure}
Figures $\ref{BOX1}$ and $\ref{BOX2}$ present boxplots of the MI-WOLS estimator across all simulation scenarios, examining complementary identification settings. Figure $\ref{BOX1}$ isolates the effects of CIT and CIO individually, whereas Figure $\ref{BOX2}$ considers the extreme cases in which both or neither assumption holds. The $x$-axis displays the four weighting schemes used in the MI-WOLS approach, and the $y$-axis reports the corresponding estimates of the target parameter $\psi_0$. Panels are arranged according to whether the mSITA assumption holds and whether either, both, or neither of the CIT and CIO assumptions is satisfied, with different colour groupings indicating different model specifications.

When mSITA is satisfied and at least one of CIT or CIO is valid, the weighted estimators using ABS, IPW, or SIPW weights yield estimates centred closely around the true effect of -2.35. This alignment is observed whenever either the treatment or the outcome model is correctly specified (i.e., in the CC, CI, and IC settings). Under these conditions, the weighted regression estimator achieves unbiased estimation even if one of the working models is misspecified, thereby demonstrating its doubly robust nature.

In contrast, the unweighted regression estimator performs well only in scenarios where the outcome model is correctly specified; otherwise, it exhibits noticeable bias. When neither CIT nor CIO holds, or when mSITA is violated, all estimators show increased deviation from the truth, with the largest bias occurring when both working models are incorrect. Overall, the patterns across the figures illustrate that the weighted approach offers double chances for consistent estimation, whereas the unweighted approach is singly robust and relies entirely on the correct specification of the outcome model.

Interestingly, the CI scenario (correct treatment model, misspecified outcome model) exhibits slightly smaller empirical variability than the both correctly specified case (CC). This behaviour reflects the fact that, when the outcome model is misspecified, estimation relies more heavily on the propensity score–based weighting, which achieves strong covariate balance and can reduce variability when correctly specified. On the other hand, when both models are correctly specified, the estimator incorporates variability from both the outcome regression and the weighting mechanism, leading to slightly larger dispersion. By comparison, the IC scenario (correct outcome model, misspecified treatment model) shows variability similar to CC, as the lack of accurate propensity score weighting limits covariate balance and prevents the same variance reduction as CI does.

\subsection{Performance of the sandwich variance estimator}
Valid statistical inference for the MI-WOLS estimator requires accurate estimation of its sampling variability. In Supplementary Materials, Section S3, we derive a robust sandwich variance estimator by expressing the MI-WOLS estimator as the solution to a system of weighted estimating equations. The resulting estimator, constructed from empirical sensitivity and variability matrices of the score functions, is robust to heteroskedasticity arising from weighting and working model misspecification \citep{freedman2006so}. In this subsection, we assess the finite-sample performance of the sandwich variance estimation through simulation.

\begin{figure}[H]
 \centering
  \includegraphics[width=1\textwidth]{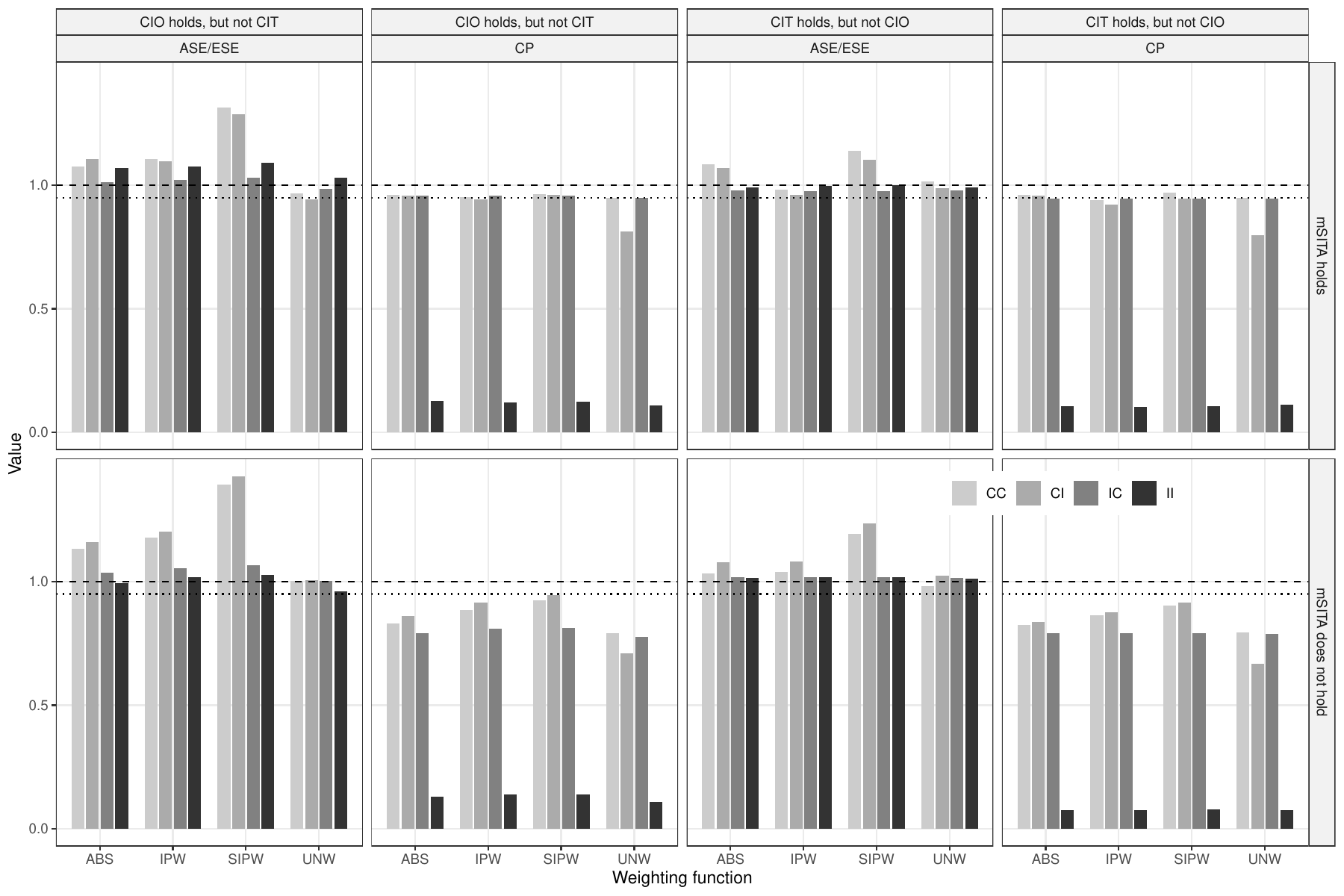}   
 \caption{Bar plots of the ratio of analytical to empirical standard errors (ASE/ESE) and 95\% coverage probabilities (CP) for the MI-WOLS estimator across different weight functions and identification scenarios. Results are shown for settings in which the mSITA assumption holds and does not hold, and for cases where either the CIT or CIO assumption holds individually.}
 \label{BAR1}
 \end{figure}
Figure~\ref{BAR1} compares the performance of the sandwich variance estimator across different identifiability assumptions, weighting schemes, and model specification scenarios. We report the ratio of the analytical standard error from the sandwich estimator representing the model-based estimate of variability to the empirical standard error reflecting the true sampling variability across simulations (ASE/ESE), along with the corresponding 95\% coverage probabilities (CP). Accurate variance estimation is indicated by ASE/ESE values close to one and coverage near the nominal level of 0.95.

When the mSITA assumption holds, and at least one of the CIT or CIO assumptions is satisfied, the sandwich variance estimator performs well across all weighting schemes. In these settings, ASE/ESE ratios are close to unity, and coverage probabilities are near nominal, even under partial model misspecification. The variance estimator remains accurate whenever at least one of the outcome or treatment models is correctly specified, reflecting the doubly robust structure of the MI-WOLS estimator. Moreover, the variance estimates remain stable even when both working models are incorrect, demonstrating the robustness of the sandwich variance estimator to model misspecification \citep{freedman2006so}. However, this robustness does not extend to CP values, which deteriorate due to bias in the effect estimates, despite the availability of reliable variance estimates. Bar plots for scenarios in which both CIT and CIO hold, or neither holds, are presented in Figure S3 of the Supplementary Materials.

\subsection{Comparison with AIPW and G-estimation}
In this subsection, we compare the finite-sample performance of the MI-WOLS estimator with two widely used doubly robust procedures: the AIPW estimator \citep{funk2011doubly,kurz2022augmented} and G-estimation \citep{robins1989analysis,vansteelandt2014structural}. These methods represent standard benchmarks for doubly robust causal inference and provide a natural reference for evaluating the behaviour of our weighted regression approach in the presence of missing confounders.

The AIPW estimator augments the inverse probability weighted estimator with an outcome regression component, thereby achieving consistency when either the treatment or the outcome model is correctly specified. For a binary treatment, the AIPW estimator of the average treatment effect is defined as: 
\begin{equation}
    \hat{\psi}_0^{(AIPW)} = n^{-1} \sum_{i=1}^{n} \left[\frac{Z_iY_i}{\widehat{\pi\left(H_i\right)}} - \frac{\left\{Z_i - \widehat{\pi\left(H_i\right)}\right\}}{\widehat{\pi\left(H_i\right)}} \widehat{m_1\left(H_i\right)}\right] - n^{-1} \sum_{i=1}^{n} \left[\frac{(1-Z_i)Y_i}{1-\widehat{\pi\left(H_i\right)}} - \frac{\left\{Z_i - \widehat{\pi\left(H_i\right)}\right\}}{1-\widehat{\pi\left(H_i\right)}} \widehat{m_0\left(H_i\right)}\right] 
\end{equation}
where $\widehat{\pi(H_i)}$ denotes the estimated propensity score and $\widehat{m_z(H_i)}$ represents the predicted (potential) outcome under treatment level $Z=z$.

G-estimation provides an alternative way to double robustness by modifying the estimating equations so that consistency holds when either the treatment or the outcome model is correctly specified. In our setting, G-estimation solves the estimating equation:
\begin{equation}
    \sum_{i=1}^{n} \left(\begin{array}{c}
         H_{i}^{\beta\top}  \\
         \left(z_i - \pi\left(H_i\right)\right) H_{i}^{\psi\top}
    \end{array}\right) \left(y_i - H_{i}^{\beta}\beta - \text{z}_iH_{i}^{\psi}\psi\right) = \bm{0}.
\end{equation}
We evaluated all estimators under the same simulation framework described in Section \ref{SUBSECSIM}. To facilitate comparison, performance was summarized using two metrics. First, we report the percentage bias relative to the estimator’s ASE (Bias/ASE (\%)). This measure reflects the magnitude of bias relative to sampling variability and provides a practical assessment of inferential validity; as noted by \citet{kang2007demystifying}, inferential performance tends to deteriorate when the bias exceeds approximately 40\% of the standard deviation. Second, we compute the relative mean squared error (MSE) of each estimator with respect to G-estimation (MSE/$\text{MSE}\_\text{GEST}$). A relative MSE greater than one indicates lower efficiency than G-estimation, whereas a value less than one indicates higher efficiency.
\begin{table}[H]
\centering
\setlength{\tabcolsep}{3pt}
\caption{Simulation study comparing the performance of the MI-WOLS estimator with AIPW and G-estimation across model misspecification and identification assumption scenarios.}
\label{COMSIMRES}
\begin{tabular}{|ll|l|rrr|rr|rrr|r|}
  \hline
  &&&\multicolumn{5}{c|}{Bias/ASE ($\%$)}&\multicolumn{4}{c|}{MSE/$\text{MSE}\_\text{GEST}$}\\ \cline{4-12}
  &&Model&\multicolumn{3}{|c|}{MI-WOLS}&&&\multicolumn{3}{|c|}{MI-WOLS}&\\ \cline{4-6} \cline{9-11}
  CIT&CIO &specification& \multicolumn{1}{|c}{ABS} & IPW & SIPW&AIPW&GEST & \multicolumn{1}{|c}{ABS} & IPW&SIPW& AIPW  \\  \hline
  & & CC & -6.06 & -6.68 & -6.37 & 21.50 & -6.36 & 1.01 & 2.19 & 2.66 & 3.23 \\ 
   $\checkmark$&$\checkmark$ & CI & -2.96 & -7.98 & -5.09 & 18.04 & -4.83 & 0.97 & 2.21 & 2.65 & 2.50 \\
    & & IC  & -4.02 & -4.38 & -4.37 & 3.37 & -4.00 & 0.98  & 0.98 & 0.98 & 0.98 \\
   & & II & 324.71 & 326.54 & 325.91 & 333.21 & 303.13 & 0.99 & 1.00 & 1.00 & 1.02 \\ \hline 
   & & CC & -3.49 & -1.33 & -1.23 & 28.69 & -3.64 & 1.05 & 2.51 & 3.19 & 2.69 \\ 
   $\times$&$\checkmark$ & CI & -8.41 & -16.96 & -13.60 & 14.19 & -9.43 & 1.06 & 2.31 & 2.82 & 2.02 \\
    & & IC  & 0.00 & -0.10 & -0.08 & -4.61 & -0.18 & 0.95 & 0.97 & 0.98 & 0.99 \\
   & & II & 304.76 & 309.73 & 305.83 & 314.85 & 294.00 & 0.99 & 1.04 & 1.04 & 1.02 \\ \hline
   & & CC & -0.89 & -0.95 & -0.59 & 25.61 & -1.43 & 1.08 & 2.29 & 2.80 & 3.48 \\ 
   $\checkmark$&$\times$ & CI & -5.12 & -13.15 & -9.49 & 12.12 & -5.86 & 1.04 & 2.19 & 2.58 & 2.56 \\
   & & IC  & 3.61 & 3.68 & 3.68 & 10.94 & 3.59 & 0.98 & 0.98 & 0.98 & 0.99 \\ 
   & & II & 321.90 & 321.98 & 321.13 & 329.14 & 319.42 & 1.00 & 1.01 & 1.02 & 1.03 \\ 
   \hline

&& CC & -73.98 & -36.97 & -33.06 & -1.95 & -78.48 & 1.07 & 2.01 & 2.54 & 2.29 \\ 
  $\times$&$\times$ & CI & -71.84 & -46.03 & -37.72 & -16.18 & -76.80 & 1.08 & 1.97 & 2.47 & 1.71 \\ 
  & & IC & -60.56 & -60.68 & -60.89& -66.16 & -57.04 & 0.94 & 0.98 & 0.98 & 1.03 \\ 
  & & II & 243.13 & 242.16 & 237.59 & 247.03 & 243.11 & 1.00 & 1.05 & 1.06 & 1.02 \\ \hline
\end{tabular}
\end{table}
Table~\ref{COMSIMRES} presents a comparison of the finite-sample performance of the MI-WOLS estimator with the AIPW estimator and G-estimation for the homogeneous treatment effect. Results are reported under settings where the mSITA assumption holds, while allowing all combinations of the CIT and CIO assumptions and model specification scenarios. All three estimators exhibit double robustness with respect to model specification, yielding valid inference whenever at least one of the treatment or outcome models is correctly specified, provided that mSITA and either CIT or CIO are satisfied. When both the CIT and CIO assumptions fail, inferential performance deteriorates substantially for all methods, as reflected by pronounced relative bias. From an efficiency perspective, the MI-WOLS estimator using ABS weights performs comparably to, and in some settings more efficiently than, G-estimation. When the treatment model is correctly specified (CC and CI configurations), G-estimation generally attains lower mean squared error than the MI-WOLS estimator based on IPW or SIPW weights, as well as the AIPW estimator. Conversely, when only the outcome model is correctly specified (IC configuration), the MI-WOLS estimator is more efficient than G-estimation. Overall, these findings indicate that the MI-WOLS approach provides competitive finite-sample performance relative to established doubly robust estimators, particularly when using ABS weights, while retaining the advantages of a flexible regression-based framework that accommodates missing confounders via balancing weights.

\subsection{Performance under treatment effect heterogeneity}
We extend the simulation study to settings with heterogeneous treatment effects by introducing effect modification through a fully observed confounder. The data-generating mechanisms are identical to those described in Section~\ref{SUBSECSIM}, except for the outcome model, which is specified as
\begin{align*}
   Y \sim Normal(1 + \psi_0 Z + \psi_1 ZC - 2.2\tau U_Y - 1.55 XR + \gamma X(1-R) + 1.8R - 1.7C + \delta_Y CR, 3).
\end{align*}
The blip function in the outcome model is given by $(\psi_0Z + \psi_1 Z C)$, where $\psi_0$ denotes the marginal treatment effect and $\psi_1$ characterizes treatment effect heterogeneity with respect to the fully observed confounder $C$. In the simulations, we set $\psi_0 = -2.35$ and $\psi_1 = -1$ while varying parameters to induce violations of mSITA, CIT, CIO, and the working model specification.

We observe that the MI-WOLS estimator performs similarly under heterogeneous treatment effects as it does under the homogeneous treatment-effect setting. Simulation results for the heterogeneous case, reported in Table S1–S4 in the Supplementary Materials, show that when the mSITA assumption holds and at least one of the CIT or CIO assumptions is satisfied, the MI-WOLS estimator consistently estimates both the main term and the effect-modification parameter, provided that either the treatment or the outcome model is correctly specified. Across these scenarios, sandwich-based standard errors closely track empirical variability and achieve near-nominal coverage for both of the blip parameters. Comparisons with G-estimation further indicate that the MI-WOLS estimator, particularly with ABS weights, often provides improved efficiency even in the presence of treatment effect heterogeneity (Tables~S5 and S6 in the Supplementary Materials).

\section{Illustration: ACEI/ARB use and kidney function in synthetic cohort data}
\label{ILLUSTRATION}
To demonstrate the practical use of the MI-WOLS estimator, we consider an application-style example motivated by the cohort study described by \citet{blake2020estimating}, which examined the relationship between use of angiotensin-converting enzyme inhibitors or angiotensin receptor blockers (ACEI/ARBs) and short-term kidney function. Because the original individual-level data are not available, we generate a synthetic dataset designed to closely reproduce the study design, covariate structure, and marginal distributions reported in the original analysis. The data are generated from a known data-generating mechanism with a true treatment effect of $-0.6831$, allowing evaluation of estimator performance in a realistic observational setting. This setting complements the simulation study by mimicking the structure of a real-world observational analysis.

The motivating study followed 570586 adult patients initiating antihypertensive therapy, with treatment defined as prescription of ACEI/ARBs and the outcome measured by kidney function within a 2 months period after treatment initiation, summarized using estimated glomerular filtration rate (eGFR). Lower values of eGFR indicate poorer renal function, and the estimand of interest is the average treatment effect comparing ACEI/ARBs with alternative antihypertensive therapies.

The simulated data include several fully observed baseline covariates capturing demographic characteristics, comorbidities, medication use, and calendar time, along with two partially observed confounders: baseline eGFR category and ethnicity. Both variables are subject to substantial missingness, creating a realistic setting in which complete-case analysis would discard a large proportion of observations.

In contrast to the illustrative analysis of \citet{blake2020estimating}, where treatment was assigned at random for demonstration purposes, we generate treatment using a stratified mechanism that depends on observed baseline characteristics. This design induces confounding and more closely reflects an observational study. The data-generating process is calibrated so that the treatment effect and marginal distributions align with those reported in the original study, providing a realistic platform for illustrating the performance of the MI-WOLS estimator in the presence of missing confounders.

In this application, the assumptions required for valid inference appear reasonable. The mSITA assumption would be violated only in the presence of unmeasured factors jointly affecting treatment assignment or the outcome and the missingness process; given that treatment is generated conditional on observed baseline characteristics, such violations are unlikely. The CIT assumption is also plausible, since unobserved confounder values are unlikely to affect prescribing decisions when such information is unavailable at the time of treatment initiation, particularly for variables such as baseline kidney function and ethnicity. In contrast, the CIO assumption is less credible because baseline kidney function remains prognostic of outcomes regardless of whether it is observed. Since validity requires mSITA and only one of CIT or CIO to hold, the MI-WOLS estimator is well justified in this setting.

\begin{table}[H]
\centering
\caption{Baseline characteristics of study participants overall and stratified by treatment status.}
\label{ILLDAEX}
\begin{tabular}{lrrrl}
  \hline
 Baseline & & \multicolumn{3}{c}{Prescribed ACEI/ARB} \\ \cline{3-5}
 characteristic& Overall & No & Yes & p-value \\ 
  \hline
n& 570586 & 414604 &155982   \\ 
  eGFR, mean (SD) &82.15 (17.82)&82.36 (17.82) &  81.58 (17.81) & $<$0.001   \\ 
  Diabetes& 83496 (14.6) &  58249 (14.0)  &  25247 (16.2)  & $<$0.001   \\ 
  Hypertension& 364614 (63.9)& 262274 (63.3)  & 102340 (65.6)  & $<$0.001   \\ 
  Cardiac failure& 31853 ( 5.6) &  21973 ( 5.3)  &   9880 ( 6.3)  & $<$0.001   \\ 
  Arrhythmia& 56200 ( 9.8)&  40638 ( 9.8)  &  15562 (10.0)  &  0.048  \\ 
  Heart disease& 118486 (20.8) &  85253 (20.6)  &  33233 (21.3)  & $<$0.001  \\ 
  Sex (Female) &298839 (52.4)& 219413 (52.9)  &  79426 (50.9)  & $<$0.001   \\ \hline 
  Age &  &  & &$<$0.001   \\ 
  $< 45$ & 342080 (60.0) & 248918 (60.0)  &  93162 (59.7)  &   \\ 
  $45-54$ & 116258 (20.4)&  83339 (20.1)  &  32919 (21.1)  &    \\ 
  $55-59$ & 47433 ( 8.3)&  34741 ( 8.4)  &  12692 ( 8.1)  &   \\ 
  $60-64$ & 34242 ( 6.0)&  25108 ( 6.1)  &   9134 ( 5.9)  &    \\ 
  $65-69$ & 19906 ( 3.5) &  14575 ( 3.5)  &   5331 ( 3.4)  &   \\ 
  $70-74$ & 7908 ( 1.4)&   5893 ( 1.4)  &   2015 ( 1.3)  &    \\ 
  $75-84$ &  2221 ( 0.4) &   1617 ( 0.4)  &    604 ( 0.4)  &    \\ 
  $\ge 85$ & 538 ( 0.1)  &    413 ( 0.1)  &    125 ( 0.1)  &     \\ \hline 
  Calendar period &  &  & & 0.285  \\ 
  $\le 2000$ & 32890 ( 5.8) &  23915 ( 5.8)  &   8975 ( 5.8)  &    \\ 
  $2001-2004$ &  85662 (15.0) &  62374 (15.0)  &  23288 (14.9)  &    \\ 
  $2005-2008$ &141807 (24.9) & 102727 (24.8)  &  39080 (25.1)  &    \\ 
  $2009-2011$ & 149111 (26.1) & 108405 (26.1)  &  40706 (26.1)  &    \\ 
  $2012-2014$ &161116 (28.2)& 117183 (28.3)  &  43933 (28.2)  &    \\ \hline
  Ethnicity&  &  &  &0.019   \\ 
  White & 217496 (38.1)& 157563 (38.0)  &  59933 (38.4)  &    \\ 
  South Asian & 7999 ( 1.4)&   5753 ( 1.4)  &   2246 ( 1.4)  &      \\ 
  Black & 5128 ( 0.9)&   3708 ( 0.9)  &   1420 ( 0.9)  &    \\ 
  Other &  2335 ( 0.4)  &   1682 ( 0.4)  &    653 ( 0.4)  &   \\ 
  Mixed &  1131 ( 0.2)  &    813 ( 0.2)  &    318 ( 0.2)  &     \\ 
  Missing &336497 (59.0) & 245085 (59.1)  &  91412 (58.6)  &    \\ \hline
  Baseline eGFR &  &  && $<$0.001   \\ 
  $< 30$ & 235571 (41.3)& 168513 (40.6)  &  67058 (43.0)  &    \\ 
  $30-44$ & 26343 ( 4.6)&  18859 ( 4.5)  &   7484 ( 4.8)  &   \\ 
  $45-59$ &  5625 ( 1.0)  &   3963 ( 1.0)  &   1662 ( 1.1)  &    \\ 
  $\ge 60$ &  1101 ( 0.2)&    747 ( 0.2)  &    354 ( 0.2)  &   \\ 
  Missing & 301946 (52.9) & 222522 (53.7)  &  79424 (50.9)  &    \\
   \hline
\end{tabular}
\end{table}
Table~\ref{ILLDAEX} summarizes baseline characteristics of the study population overall and stratified by ACEI/ARB prescription status. Here $p$-values correspond to comparisons between treatment groups, assessing differences in baseline characteristics (e.g., t-tests for continuous variables and chi-square tests for categorical variables). The cohort is predominantly female and relatively young, with hypertension being the most common comorbidity, while diabetes and other cardiovascular conditions are less prevalent. Most baseline characteristics, including comorbidities, sex, age, ethnicity, and baseline eGFR, differ significantly between treated and untreated groups, suggesting the presence of potential confounding. Substantial missingness is observed for ethnicity (59\%) and baseline eGFR (52.9\%), affecting more than half of participants; among those with observed values, patients are predominantly White and commonly have moderate to severe reductions in kidney function.
\begin{table}[H]
\centering
\caption{Estimated treatment effects and 95\% confidence limits (CL) from simulated observational data comparing ACEI/ARB prescription at baseline versus no prescription, under varying treatment ($\pi$-) model and outcome ($y$-) specifications; the true treatment effect is $-0.6831$.}
\label{ILLANARES}
\begin{tabular}{lccccrc}
  \hline
Method&$\pi$-model & $y$-model & Estimate&Bias&SE & 95\% CL\\ 
  \hline
\multicolumn{2}{l}{MI-WOLS estimator}&&&&&\\ 
 \multicolumn{1}{r}{UNW} &$-$&$\checkmark$& -0.706 & -0.022 & 0.044 & [-0.791, -0.620] \\ 
  &$-$&$\times$& -0.825 & -0.142 & 0.051 & [-0.925, -0.725]  \\ \cline{2-7}

  \multicolumn{1}{r}{ABS} &$\checkmark$&$\checkmark$& -0.705 & -0.022 & 0.044 & [-0.791, -0.620] \\
  &$\checkmark$&$\times$& -0.707 & -0.024 & 0.051 & [-0.807, -0.606]\\ 
  &$\times$&$\checkmark$& -0.706 & -0.022 & 0.044 & [-0.791, -0.620] \\ 
  &$\times$&$\times$& -0.825 & -0.142 & 0.051 & [-0.925, -0.725]\\ \cline{2-7}

  \multicolumn{1}{r}{IPW} &$\checkmark$&$\checkmark$& -0.708 & -0.025 & 0.044 & [-0.793, -0.623] \\
  &$\checkmark$&$\times$& -0.709 & -0.026 & 0.051 & [-0.810, -0.608] \\ 
  &$\times$&$\checkmark$& -0.708 & -0.025 & 0.044 & [-0.793, -0.622] \\ 
  &$\times$&$\times$& -0.828 & -0.145 & 0.051 & [-0.929, -0.728] \\ \cline{2-7}

  \multicolumn{1}{r}{SIPW} &$\checkmark$&$\checkmark$& -0.708 & -0.025 & 0.044 & [-0.793, -0.623] \\
  &$\checkmark$&$\times$& -0.709 & -0.026 & 0.051 & [-0.810, -0.608] \\ 
  &$\times$&$\checkmark$& -0.708 & -0.025 & 0.044 & [-0.793, -0.623] \\ 
  &$\times$&$\times$& -0.828 & -0.145 & 0.051 & [-0.929, -0.728] \\ \hline

  \multicolumn{2}{l}{Doubly robust competitors}&&&&&\\ 
   \multicolumn{1}{r}{AIPW} &$\checkmark$&$\checkmark$& -0.708 & -0.025 & 0.044 & [-0.793, -0.623] \\ 
  &$\checkmark$&$\times$& -0.709 & -0.026 & 0.051 & [-0.810, -0.608] \\ 
  &$\times$&$\checkmark$& -0.708 & -0.025 & 0.043 & [-0.793, -0.623]\\ 
  &$\times$&$\times$& -0.828 & -0.145 & 0.051 & [-0.929, -0.728] \\ \cline{2-7}

     \multicolumn{1}{r}{G-estimation} &$\checkmark$&$\checkmark$& -0.705 & -0.022 & 0.044 & [-0.791, -0.620] \\
  &$\checkmark$&$\times$& -0.707 & -0.024 & 0.044 & [-0.792, -0.621] \\ 
  &$\times$&$\checkmark$& -0.706 & -0.022 & 0.044 & [-0.791, -0.620] \\ 
  &$\times$&$\times$& -0.825 & -0.142 & 0.051 & [-0.925, -0.725] \\ \hline
\end{tabular}
\end{table}
Table~\ref{ILLANARES} summarizes the estimated treatment effects, bias, standard errors (SEs), and 95\% confidence limits (CLs) obtained from the simulated observational data under various model specification scenarios. The MI-WOLS estimator provides unbiased effect estimates when either the treatment model or the outcome model is correctly specified, demonstrating the expected double robustness. In contrast, misspecification of both models results in increased bias and reduced inferential accuracy. The unweighted missing-indicator regression approach yields unbiased estimates only when the outcome model is correctly specified, but exhibits noticeable bias when the model is misspecified. Similarly, the AIPW and G-estimation methods exhibit comparable robustness, producing reliable estimates when at least one of the working models is correctly specified. Given the large sample size, differences in SEs among the doubly robust estimators are minimal, resulting in largely similar performance across methods. Overall, the results indicate that ACEI/ARB use is associated with an approximately 0.7-unit reduction in kidney function, and this association is statistically significant, as none of the corresponding 95\% CLs include zero. Nevertheless, the findings presented here are intended to illustrate the methodological performance of the proposed approach and should not be used as a basis for substantive clinical conclusions. Consistent with the simulation studies, these results further demonstrate the double robustness of the proposed method and its practical reliability in settings that mimic real-world observational analyses. Together, these findings highlight the value of extending balancing-weight approaches to settings with partially observed confounders, providing a principled and practically robust framework for causal inference in the presence of incomplete data.

\section{Discussion}
\label{DISCUSSION}
This article extends the missing–indicator framework for causal inference with partially observed confounders by developing a doubly robust MI-WOLS estimator. Existing approaches developed by \citet{blake2020propensity,blake2020estimating} include a singly robust propensity score–based estimator and a singly robust outcome regression estimator. Both rely on the mSITA assumption together with at least one of the CIT or CIO assumptions, but each requires the correct specification of a single working model. Consequently, inference may be sensitive to model misspecification when the validity of the chosen model is uncertain.

Our primary contribution is to unify these two approaches within a weighted regression framework that achieves double robustness. Building on the balancing-weight principles from the DTR literature, especially dWOLS, we extend its core ideas to settings with partially observed confounders. In contrast to existing balancing-weight approaches that rely on fully observed covariates, we incorporate missingness indicators directly into the balancing conditions. This yields a principled extension of balancing-weight methods to the missing-indicator setting, under which the estimator remains consistent when either the treatment or outcome model is correctly specified, provided that mSITA holds together with either CIT or CIO. The resulting approach retains the interpretability and simplicity of regression-based estimation while providing a `double chance' for valid inference.

The simulation results support the theoretical properties of the estimator. When mSITA holds and at least one of CIT or CIO is satisfied, the MI-WOLS estimator exhibits negligible bias and near-nominal coverage whenever either the treatment or the outcome model is correctly specified, while performance deteriorates when both models are misspecified or when identification assumptions fail. Compared with AIPW and G-estimation, MI-WOLS demonstrates competitive finite-sample performance, especially when ABS weights are used. Although AIPW is doubly robust, it is designed to estimate average treatment effects at the population level and does not directly inform individualized treatment decisions. G-estimation, on the other hand, supports individualized decision-making and also affords double robustness, but its implementation can be challenging in practice, especially for less experienced analysts. The proposed MI-WOLS estimator maintains the double robustness of G-estimation while offering the practical advantages of a weighted regression framework. 

From a practical perspective, the MI-WOLS estimator can be readily implemented using standard regression software and produces interpretable estimates, making it accessible to applied researchers. At the same time, it inherits the fundamental limitations of missing-indicator approaches. In particular, it is designed to address missingness in confounders only and does not accommodate missing treatment or outcome data. Bias will arise if the mSITA assumption is violated or if both the CIT and CIO conditions fail, and efficiency can be reduced when extreme propensity scores are present. Future work may extend this framework to more complex settings, including longitudinal treatments, noncontinuous outcomes, and the use of flexible machine learning methods for nuisance parameter estimation. Overall, this study shows that integrating balancing-weight regression with the missing-indicator framework provides a flexible, theoretically grounded, and practically robust approach to causal inference with partially observed confounders.

\bibliography{References}
\bibliographystyle{apalike}

\end{document}